\documentclass[11pt,draftcls,onecolumn]{IEEEtran}
\usepackage{graphicx}
\usepackage{caption}
\usepackage{subfigure}
\usepackage{amsmath}
\usepackage{amssymb}
\usepackage{bm}
\usepackage{color}
\usepackage{url}

\newcommand{\tS}{{\tt S}}

\newcommand{\tD}{{\tt D}}

\newcommand{\tE}{{\tt E}}

\graphicspath{{./fig/}}

\def\bl#1{{#1}}

\begin{document}
\title{Rethinking the Role of  Interference in Wireless Networks}
 \markboth{\textit{A Manuscript Accepted in  The IEEE     Communications Magazine} }{}

\author{Gan Zheng, Ioannis Krikidis, Christos Masouros, Stelios Timotheou,  Dimitris-Alexandros
Toumpakaris and Zhiguo Ding
\thanks{Gan Zheng is with School of Computer Science and Electronic Engineering, University of Essex, UK, E-mail: {\sf ganzheng@essex.ac.uk.}
 He is also affiliated with Interdisciplinary Centre for Security, Reliability and Trust (SnT),
  University of Luxembourg, Luxembourg.}
\thanks{Ioannis Krikidis and Stelios Timotheou are with the Department of Electrical and Computer Engineering, University of Cyprus, Cyprus.
E-mail: {\sf \{krikidis, timotheou.stelios\}@ucy.ac.cy}.}
\thanks{Christos Masouros is with the Department of Electronic and Electrical Engineering, University College London, UK. Email: {\sf c.masouros@ucl.ac.uk}.}
\thanks{Dimitris-Alexandros Toumpakaris is with the Department of Electrical \& Computer Engineering, University of Patras, Greece. Email: {\sf dtouba@upatras.gr}.}
\thanks{Zhiguo Ding is with the School of Electrical, Electronic, and Computer Engineering,  Newcastle University
Newcastle, NE1 7RU, UK. Email: {\sf zhiguo.ding@newcastle.ac.uk}.}
 }

\date{\today}
 \maketitle

\begin{abstract}
This article re-examines the fundamental notion of interference in
wireless networks by  contrasting traditional approaches to new
concepts that handle interference in a creative way.
Specifically, we discuss the fundamental limits of the interference
channel  and present the interference alignment technique and its
extension of signal alignment techniques. Contrary to this
traditional view, which   treats interference as a detrimental
phenomenon,  we introduce three concepts that handle interference as
a useful resource. The first concept exploits interference at the
modulation level and leads to simple multiuser downlink precoding
that provides significant energy savings. The second concept uses
radio frequency radiation for energy harvesting and handles
interference as  a source of green energy. The last concept refers
to a secrecy environment and uses interference as an efficient means
to jam potential eavesdroppers.  These three techniques bring a new
vision about   interference in wireless networks and motivate a
plethora of potential new applications and services.
\end{abstract}

\section{Introduction}
 Resources (e.g. time, frequency, code) have to be shared by
 multiple users in wireless networks. Therefore,  interference has long been
 considered as a deleterious  factor that  limits the system capacity. In   conventional
 communications systems,  the design objective is to allow users to share a medium with minimum or no
 interference. Thus, great efforts are made to avoid, mitigate and cancel interference. For instance, to support multiple users, orthogonal access
methods in   time, frequency, code as well as spatial domains
have been used in different generations of cellular systems. In
future-generation heterogeneous cellular networks, due to the
increasing number of uncoordinated low-power nodes in such as
femtocells to improve the coverage and capacity, interferences need
to be mitigated in multiple domains, rendering their management a
challenge.

Interference mitigation/avoidance techniques provide convenient
mechanisms to allow multiple users to share the wireless medium.
However, they lead to   inefficient use of wireless resources. One
may ask whether to cancel or mitigate interference is always the
optimal way of utilizing wireless resources. Indeed, there has been
growing interest in exploiting interference to improve the
achievable rate, the reliability and the security of wireless
systems. Recently, new views on interference have resulted in
advanced interference-aware techniques, which, instead of mitigating
interference, explore the potential of using interference. We
present two examples from the literature to illustrate the ideas.

In his early work of  dirty paper coding \cite{Costa},  Costa proved
the striking result that  interference known at the transmitter but
not at the receiver does not affect the capacity of the Gaussian
channel. The optimal strategy to achieve this interference-free
capacity is to code along interference, while cancelling
interference is strictly sub-optimal.  Another example is
coordinated multipoint  or multi-cell coordination,  where base
stations (BSs) cooperate to serve their own and out-of-cell users.
In the downlink, the cooperating
BSs work together to jointly optimize the transmitter strategies
such as power, time and beamforming design to control the inter-cell
interference. Cell-edge users who suffer most from the inter-cell
interference now benefit most from this coordination. In the uplink,
jointly decoding is performed in BSs, so signals from users in other
cells are no longer treated as interference, but as useful signals.

The purpose of this article is to re-examine the notion of
interference in communications networks and introduce a new paradigm
that considers interference as a useful resource. We first give an
overview from the information theoretic standpoint as a
justification for rethinking the role of interference in wireless
networks.   We then introduce interference alignment and signal
alignment as effective means to handle interference and increase the
achievable rates. Departing from this traditional view, we present
three novel techniques of interference exploitation that aim to
improve the performance of wireless networks. The first technique is
a data-aided precoding scheme in the multiuser downlink that
judiciously makes use of the interference among users as a source of
useful signal energy. In the second technique, we consider
simultaneous information and energy transfer; in such a system,
while interference links are harmful for information decoding, they
are useful for energy harvesting. Thus, a favorable tradeoff is
demonstrated. The third technique leverages interference in physical
layer secrecy as an effective way to degrade the channel of the
eavesdropper and increase the system's secrecy rate.

\section{Interference from the information theoretic standpoint}
\begin{figure}[h]
\centering
\includegraphics[width=4in]{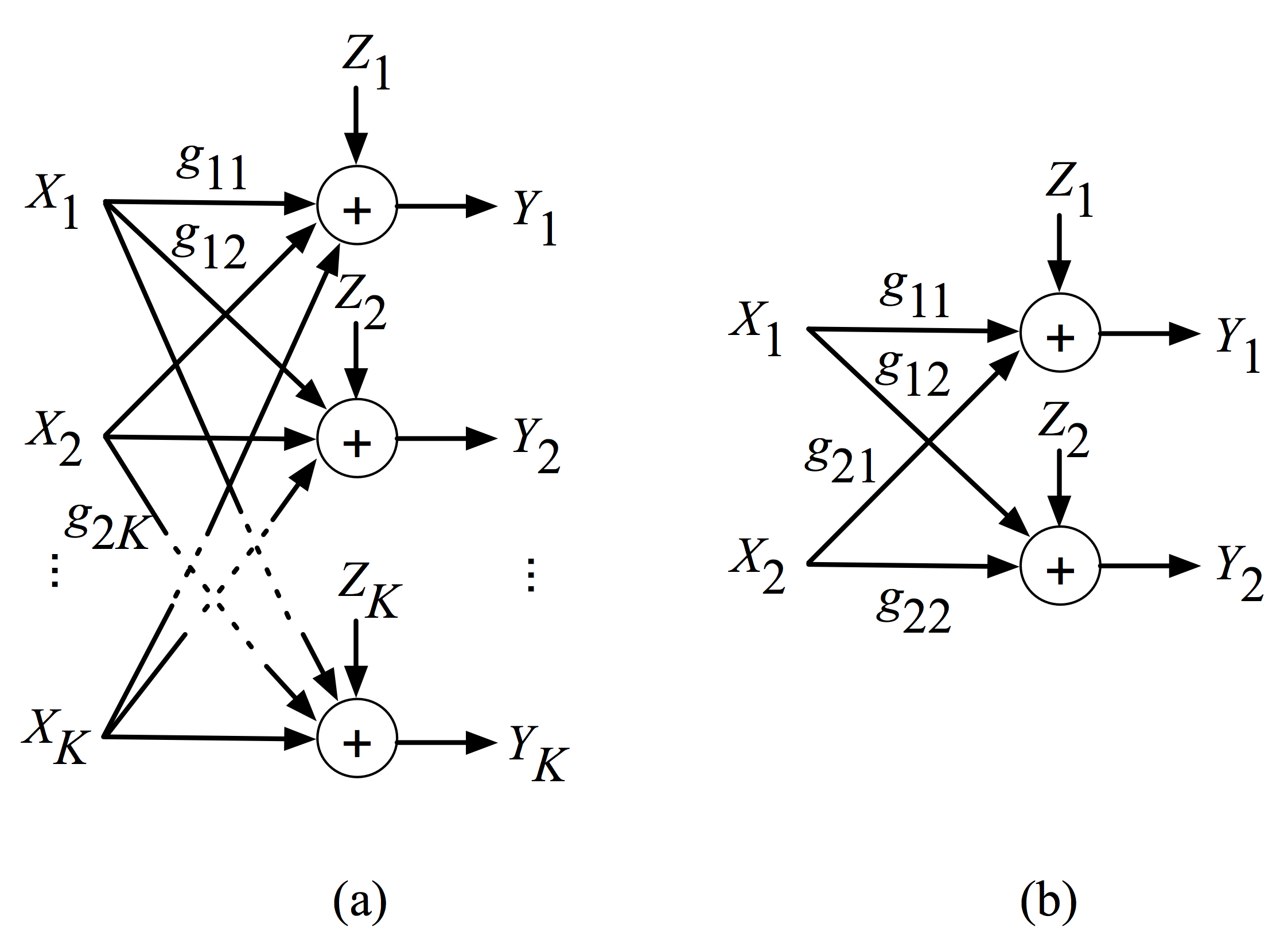}\\
\caption{(a) The $K$-user Gaussian Interference Channel (G-IC) (b)
The 2-user G-IC.}\label{fig:gic}
\end{figure}
We first present an overview of results on interference from
information theory. The Interference Channel (IC) models
simultaneous transmission by non-cooperating transmitters and
receivers.  The messages of each link are encoded only by the
corresponding transmitter, and the receiver does not have access to
the signals of other receivers. Figure~\ref{fig:gic}(a) depicts the
$K$-user Gaussian Interference Channel (G-IC). Each of $K$
transmitters wants to send a message to the corresponding receiver.
Receiver $k$ bases its decision on signal $Y_k$, which contains not
only the (scaled) useful signal $X_k$, but also interference and
Gaussian noise.

Despite its apparent simplicity, to this date it is not known what
the optimal way of transmitting over the G-IC is,  not even for the
2-user G-IC shown in Fig.~\ref{fig:gic}(b). Nevertheless,
significant progress has been made recently, and results from
information theory have started influencing the design of wireless
networks. {  The optimal decoding strategy depends on the power of
interference compared to the direct links. Interference should be
treated as noise when it is very weak. The exact conditions for the
2-user G-IC to be in the \emph{very weak} regime can be found in
\cite{ElGamal_Kim}. In information theoretic terms, the messages of
both transmitters are \emph{private}, since they are only decoded at
the intended receiver. On the other hand,  when the power of the
interfering signal exceeds the power of the signal of interest
(\emph{strong} interference), the optimal strategy is to also decode
interference at the receivers. In this case, both messages are
\emph{public}. If the power of the interference exceeds an even
higher threshold the G-IC is in the \emph{very strong} interference
regime and the rate that can be achieved by each link is the same as
if the interferer did not exist, \emph{i.e.,} interference does not
impact the achievable rates. Nevertheless, the receiver does need to
decode interference in addition to the signal of interest.  Clearly,
there are costs associated with interference-aware decoding. The
receivers are more complex, synchronization  is essential, and each
receiver needs to estimate not only its own channel, but also the
cross-channel coefficients.}

 { The most challenging situation
arises when the power of the interference  is of the order of the
power of the signal of interest. To this date it is not known what
the best way to transmit and decode is. A strategy that combines
public and private messages (the so-called Han \& Kobayashi scheme
\cite{ElGamal_Kim})  achieves higher rates compared to treating
interference as noise or avoiding it via orthogonal transmission, or
attempting to decode all messages at each receiver.} Moreover, it
has been shown that as the {\bl signal-to-noise ratio }(SNR) grows
to infinity, a simplified Han \& Kobayashi scheme can attain the
capacity region within $1/2$ bit \cite{ElGamal_Kim}. In addition to
providing evidence that strategies based on the Han \& Kobayashi
scheme may be the best for the 2-user G-IC, this result may prove
useful in future wireless networks with small cell size that will
operate at high SNRs and will therefore be limited by interference
rather than by noise.

Devising good strategies for the $K$-user G-IC seems to be even more
challenging, and the Han \& Kobayashi scheme does not appear to
extend to the $K$-user G-IC in a straightforward manner. A promising
direction towards { finding} good strategies for the $K$-user G-IC
appears to be dealing with the combined interference by all $K-1$
users at each receiver instead of decoding separately the
interference by each user. {  Furthermore, a deterministic
approximation framework has been developed for the G-IC, which
enables the construction of structured codes \cite{ElGamal_Kim}. By
employing structured lattice codes, which are also used in other
scenarios, such as multi-way relay channels, it is possible to
attain the capacity region of the G-IC within a constant gap
\cite{NG11}.   Very recently,  there has been an interesting finding
that connects   topological interference management and index coding
\cite{Jafar14}.  This connection can be leveraged to calculate rate
regions that are within a constant gap from capacity and to develop
transmission schemes over wireless networks.
 The existing index coding solutions are then translated to interference
 management solutions via a family of elegant achievability schemes of  interference alignment (IA) that has generated
significant interest, which is discussed in more detail in Section
\ref{sec:ia}. }

System designs that operate based on the best known achievability
schemes of information theory being the ultimate goal, in the
meantime improvements in performance can also be attained by
incorporating interference-aware schemes in current systems. In
\cite{LTY11} it is shown that when the transmitters use discrete
constellations and interference-aware detectors are employed at the
receivers the achievable rates over the fading G-IC are limited by
the SNR rather than by the signal-to-interference-plus-noise ratio
(SINR).

\section{Interference and Signal  Alignment}
\label{sec:ia}  Prior to the invention of  IA  \cite{Jafar-ia},
interference avoidance has been achieved by relying on the use of
orthogonal frequency or time channels. And when interference is
inevitable, conventional approaches are to adopt advanced
decoding/detecting algorithms by treating interference as noise.

The success of IA lies in the fact that it efficiently exploits the
rich  degrees of freedom available from the time/frequency/spatial
domains. By a careful coordination among the transmitters, the use
of IA can ensure that all the interference is aligned together to
occupy   one half of the signal space at each receiver, leaving the
other half available to the desired signal.  As a result, the
per-user rate achieved by IA for the interference channel with $K$ {
pairs of single-antenna  transceivers } is $C(SNR)=\frac{1}{2}\log
(SNR)+o(\log(SNR))$. This  result is surprising since a traditional
view is that such a $K$-user scenario is interference limited and
hence the per-user rate is diminishing by increasing the number of
users. As a result, the use of  IA ensures that   the spectral
efficiency of wireless communications can be improved significantly
since more users sharing the same bandwidth yields a larger system
throughput.

In addition to interference channel, the concept of IA has also been
applied to other communication scenarios, including multiple access
channel, broadcast channel, one/two-way relaying channel as well as
physical layer security.    In practice, the implementation of IA is
not trivial since the global channel state information   at   each
transmitter (CSIT) is required, which is challenging particularly
for the case with fast  time varying channels. Two types of
approaches to realize IA in practice have been proposed.

{ One is to apply advanced feedback techniques and existing results
have demonstrated that the number of fedback bits needs to be
proportional to the {\bl SNR}  in order to achieve nearly optimal
performance \cite{Krishnamachari-13}.} The other is to exploit the
coherent structure of channels and apply manipulations analog to
space time coding at the transmitters. As a result, the concept of
IA can be implemented even when the channel information is not
available to the transmitters.

\begin{figure}[h]
\centering
\includegraphics[width=6.5in]{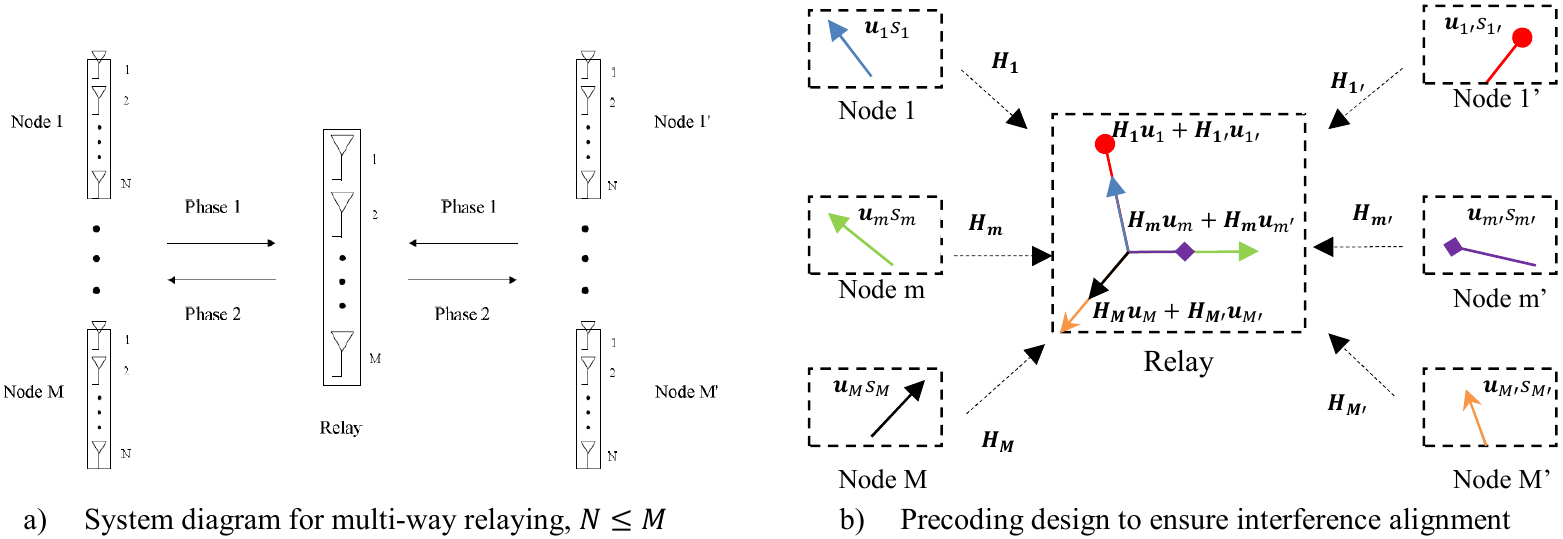}\\
\caption{Illustration of the concept of signal alignment.}
\label{fig:ia}
\end{figure}

The concept of signal alignment can be viewed as  an   extension of
IA in the context of bi-directional communications \cite{Lee_lim10}
and \cite{ZDing_Poor13}. For example, consider a multi-pair two-way
relaying communication scenario as shown in Fig. \ref{fig:ia}, where
$M$ pairs of source nodes exchange information with their partners
via the relay. Each source node is equipped with $N$ antennas, and
the relay has $M$ antennas.  As can be seen from  Fig. \ref{fig:ia},
the relay observes $2M$ incoming signal streams, and needs to have
at least $2M$ antennas in order to separate these signals. The use
of signal alignment is to effectively suppress  intra-pair
interference and reduce the requirement to the number of antennas at
the relay. Particularly, by carefully designing the precoding
vectors at the sources, the intra-pair interference is aligned
  at the relay, which means that the original $2M$ signal
streams are merged into $M$ streams. As a result, a relay with only
$M$ antennas can accommodate $2M$ incoming signals, which is
particularly important for practical scenarios where nodes are
equipped with a limited number of antennas. At the user end, each
receiver can first subtract its own information, the so-called
self-interference, and then detect the information from its partner,
a way analogous to network coding.

\section{Data-Aware Interference Exploitation for Multiuser Transmission}
\label{sec:interference_exploitation} The \textit{a priori}
knowledge of interference is readily available at downlink
transmission, where   CSIT  combined with the knowledge of all data
symbols intended for transmission can be used to explicitly predict
the resulting interference between the symbols. Despite the insights
in \cite{Costa}, the majority of existing precoding implementations
attempt to eliminate, cancel or pre-subtract interference. Only
recently however, has there been a rising interest in making use of
the interference power to enhance the useful signal \cite{cchristos,
CI_ComMag}. Indeed, it has been shown that interference can
contribute constructively to the detection of the useful signal and
this phenomenon can be utilized in the CSIT-assisted downlink
transmission and other known-interference scenarios to improve
performance without raising the {\bl transmit} power.

\begin{figure}[h]
  \centering
  \includegraphics[width=0.8\textwidth]{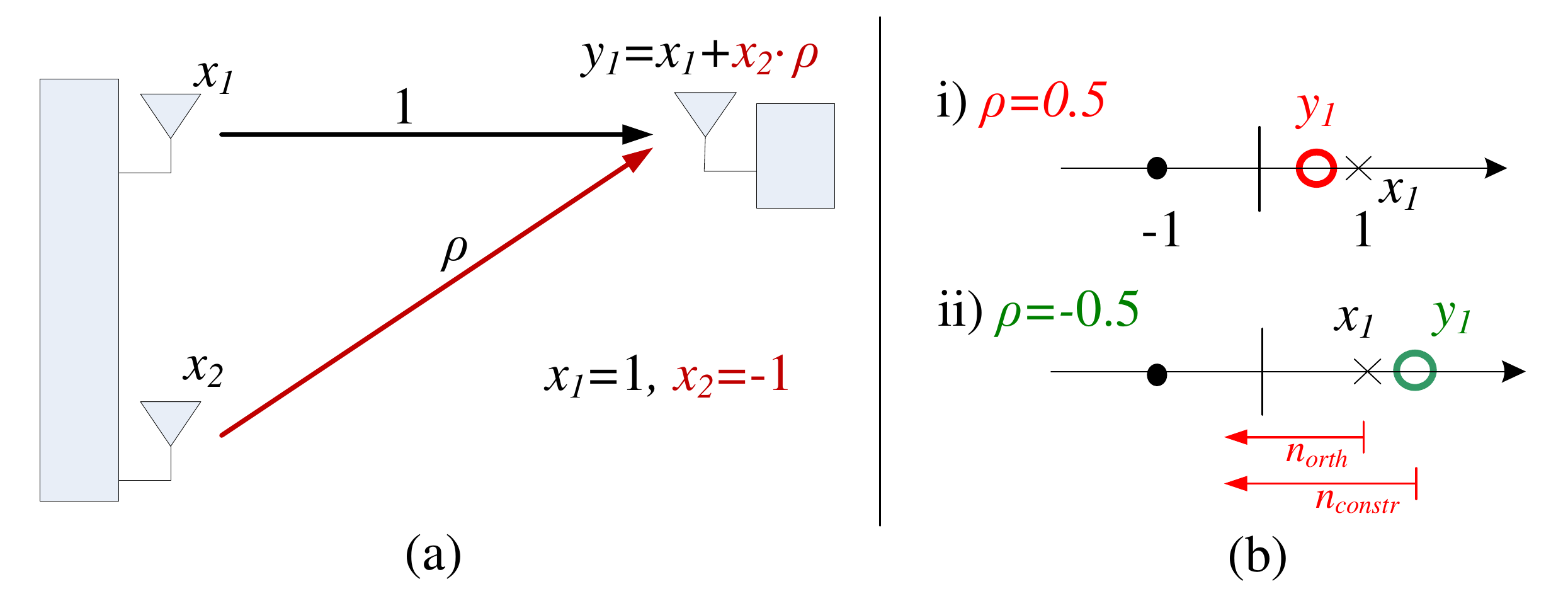}\\
  \caption{The concept of constructive interference --- a two-user example.}\label{Interf}
\end{figure}

  To clarify the above fundamental concept, a trivial example of two users is shown in Fig. \ref{Interf}(a), where
  we define the desired symbol as $x_1$ and the interfering symbol as $x_2$. Without loss of generality we assume that these belong to a
  Binary Phase Shift Keying (BPSK) constellation and that $x_1=1, x_2=-1$. For illustration purposes, we assume a lossless channel
  from the intended transmitter to the receiver and an interfering channel represented by the coefficient $\rho$. Ignoring noise, the received signal is
\begin{equation} \label{y1}
y_1=x_1+x_2\cdot \rho,
\end{equation}
where $x_2\cdot \rho$ is the interference. Note that this model also
corresponds to a multi-antenna transmission with matched filtering
where the correlation between the two channels is $\rho$. In Fig.
\ref{Interf}(b) two distinct cases are shown, depicting the
transmitted ($\times$) and received (o) symbols for user 1 on the
BPSK constellation. In case i) with $\rho=0.5$ it can be seen from
\eqref{y1} that $y_1=0.5$. The destructive interference from user 2
has caused the received symbol of user 1 to move towards the
decision threshold (imaginary axis). The received power of user 1
has been reduced and its detection is prone to low-power noise. In
case ii) however, for $\rho=-0.5$ \eqref{y1} yields $y_1=1.5$, and
hence the interference   is constructive. The power received has
been augmented due to the interference from user 2 and now its
detection is tolerant to higher noise power ($n_{constr}$ compared
to $n_{orth}$). It should be stressed that in both cases the
transmit power for each user is equal to one. Note that, while the
above example refers to a two-user transmission scenario for
illustration purposes, the fundamental concept can be extended to
more users, multipath transmission, inter-cell interference  and
other generic interference-limited systems.

\begin{figure}[h]
  \centering
   \includegraphics[width=0.6\textwidth]{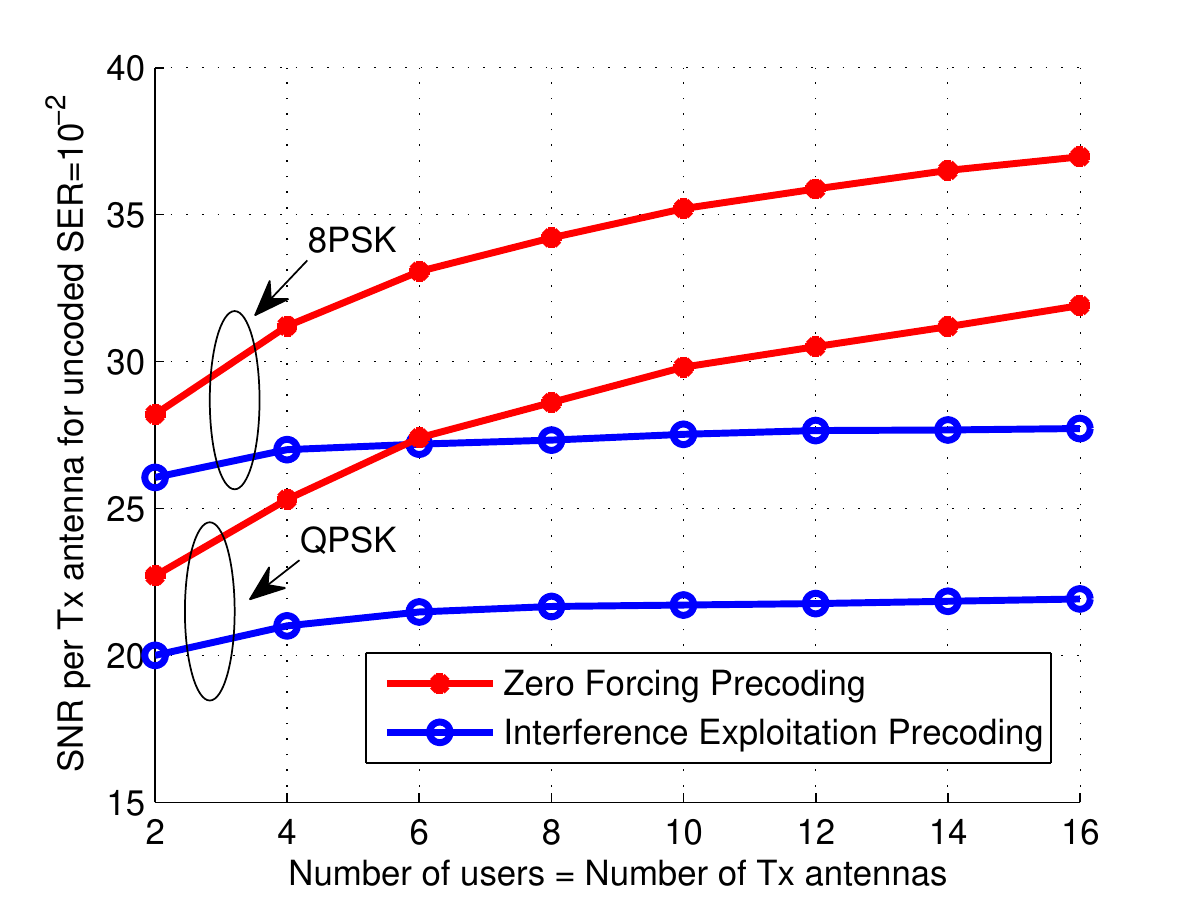}\\
  \caption{Required SNR per transmit antenna for an uncoded SER of $10^{-2}$ with increasing numbers of users and transmit antennas.}\label{RES}
\end{figure}

 Clearly, there are critical gains to be drawn from the
exploitation of constructive interference in interference-limited
transmission. As a first step, analytical constellation-dependent
characterization criteria for systematically classifying
interference to constructive and destructive have been derived in
\cite{cchristos, CI_ComMag} and references therein for PSK
modulation. {\bl Early work carried out on a simple linear precoding
technique has reported multi-fold increase in received SNR for
fixed transmit power compared to zero-forcing (ZF) beamforming
\cite{cchristos}. This can be nontrivially translated to multi-fold
savings in transmit power for a fixed received SNR. } A
representative result is shown in Fig. \ref{RES}(a) where the
required  SNR  per transmit antenna in a cellular downlink for an
uncoded symbol error rate (SER) of $10^{-2}$ is shown for increasing
numbers of single-antenna users. The results compare the widely
known  ZF  precoding with the interference exploitation precoding of
\cite{cchristos} for QPSK and 8PSK modulation. Significant SNR gains
of up to 10dB (a 10-fold transmit power reduction) can be observed
between the two techniques, by simply exploiting the existing
constructive interference.

Further work has investigated the application of this concept on
advanced nonlinear precoding, yielding further significant gains in
the transmit power. More recent work has extended this concept to
inter-cell interference exploitation in multi-cellular transmission
scenarios \cite{CI_ComMag}. The important feature in all the above
techniques is that the performance benefits are drawn not by
increasing the {\bl transmit} power of the useful signal, but rather
by reusing interference power that already exists in the
communication system; a source of green signal power that with
conventional interference cancellation techniques is left
unexploited.

\section{Wireless information and energy transfer}\label{sec:eh}
Energy harvesting (EH) communication systems that can scavenge
energy from a variety of natural sources (solar, wind, etc.) for
sustainable network operation have attracted significant interest.
The main limitation of conventional EH sources is that they are
weather-dependent and thus not always available.
A promising
harvesting technology that could overcome this bottleneck is radio
frequency (RF) energy transfer where the ambient RF radiation is
captured by the receiver antennas and converted into a direct
current   voltage through appropriate circuits (rectennas). The
concept of RF-EH is not new; over 100 years ago, Nicola Tesla proved
and experimentally demonstrated the capability of transferring
energy wirelessly.  The integration of RF-EH technology into
communications networks opens new challenges in the analysis and
design of transmission schemes and protocols. Multi-user
interference, which is the main degradation factor in conventional
networks, can be viewed as useful energy signals that could be
exploited for harvesting purposes. Although from an information
theoretic standpoint the same signal can be used for both decoding
and EH, due to practical hardware constraints, simultaneous energy
and information transmission is not possible with existing rectenna
technology. Two practical receiver approaches for simultaneous
wireless power and information transfer are a) ``time switching''
(TS), where the receiver switches between decoding information and
harvesting energy and b) ``power splitting'' (PS), where the
receiver splits the received signal in two parts for decoding
information and harvesting energy, respectively \cite{Zhang_PS_TS}.

An interesting implication of the PS technique is that in multiuser
networks harvested energy at a particular receiver can emanate
either from sources that intentionally transmit towards that
direction or from other sources whose signal is traditionally
perceived by that receiver as interference. Nonetheless, in this
case the contribution of useful and interfering signals towards the
satisfaction of any RF-EH requirements is equally important. This
implication changes completely the design philosophy of such
networks, as interference becomes useful.

This concept was demonstrated for the multiple-input single-output
(MISO) interference channel where $K$ transmitters, each one with
$K$ antennas, communicate with $K$ single-antenna receivers
\cite{Krikidis_ICC}; each receiver is characterized by both
quality-of-service (QoS) and {\bl RF-EH} constraints, while PS is
used for simultaneous information/energy transfer. The QoS
constraint requires the  SINR to be higher than {\bl a \bl given
threshold}, while the {\bl RF-EH} constraint requires the power
input to the RF-EH circuitry to be above a threshold. In this
framework, an interesting non-convex optimization problem arises in
selecting the beamforming weights and the power of the transmitters
as well as the power splitting ratios at the receivers so as to
minimize the total {\bl transmit} power. {The problem can be solved
optimally using semidefinite programming, while  traditional
beamformers can be employed to obtain suboptimal but low-complexity
solutions.} An interesting conclusion, is that for ZF beamforming
there always exists a unique, optimal, closed-form power allocation.

\begin{figure}[h]
  \centering
  \includegraphics[width=0.7\textwidth]{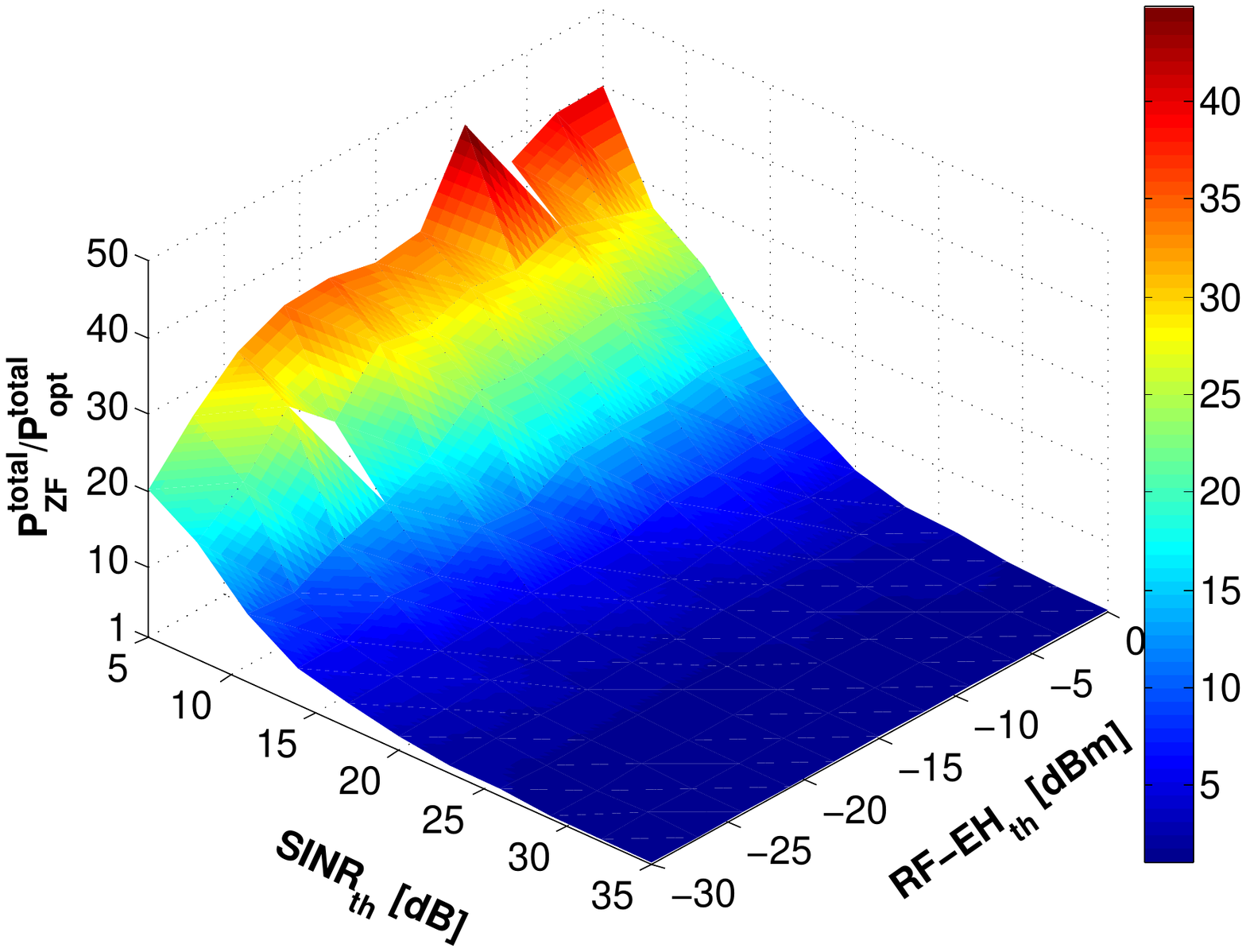}\\
  \caption{Transmitted power benefit from optimal exploitation of interference compared to ZF beamforming to achieve SINR and RF-EH constraints in   MISO interference channel.}
  \label{EH}
\end{figure}

The benefit of exploiting interference in the context of RF-EH is
illustrated in Fig. \ref{EH}, which depicts the {\bl transmit} power
ratio between  ZF and optimal beamforming for varying SINR and RF-EH
thresholds ($K=8$). The figure indicates that by exploiting
interference the {\bl transmit}  power can be significantly reduced,
especially for low SINR. The reason is that for low SINR there is
room to increase interference which is beneficial for RF-EH. In
contrast, high SINR thresholds requires almost full cancellation of
interference; hence, the solutions obtained from ZF are almost
optimal. The benefits of interference exploitation can also be seen
with respect to the RF-EH constraints: when the RF-EH threshold
increases, the ZF/optimal power ratio increases because the optimal
scheme manages interference better. However, the effect of the SINR
constraint on the {\bl transmit}  power ratio is more significant
compared to the RF-EH constraint.

\section{Interference-aided Secrecy Rate Improvement}
\label{sec:security}

Due to the growing wireless applications, confidentiality and secret
transmission has become an increasingly important issue. Recently,
securing wireless communications at the physical (PHY) layer has
been studied as a complimentary measure to upper layer cryptographic
techniques.  In the presence of eavesdroppers who passively overhear
the communication, intentional interference plays a key role to
improve the secrecy rate. This is understandable since interference
will affect both systems, however, if properly designed, it can be
an advantage for the legitimate system. {\bl This is indeed true  as
it has been shown  in \cite{Interference_security} that, the
exploration of aggregated interference together with location and
channel quality information, can significant improve network
secrecy. In the following, we review several approaches that utilize
interference to confuse the eavesdropper in a simple point-to-point
network.}

Consider a basic 3-node system which consists of a source $\tS$, a
destination $\tD$ and an eavesdropper $\tE$.  When $\tS$ has
multiple antennas, it can    transmit information bearing signal to
$\tD$ in the range space of the channel to $\tD$ and also generate
artificial noise (AN) to $\tE$ in its null space simultaneously. In
this way, even without knowledge of the instantaneous CSI of the
eavesdropper, the generated AN does not interfere with the
legitimate receiver $\tD$ and only affects the eavesdropper node
$\tE$. The same principle applies if there are trusted helper relays
who could form distributed beamforming to transmit  cooperative
jamming signals to $\tE$.

\begin{figure}[h]
\centering
\begin{subfigure}
\centering
\includegraphics[width=0.45\textwidth]{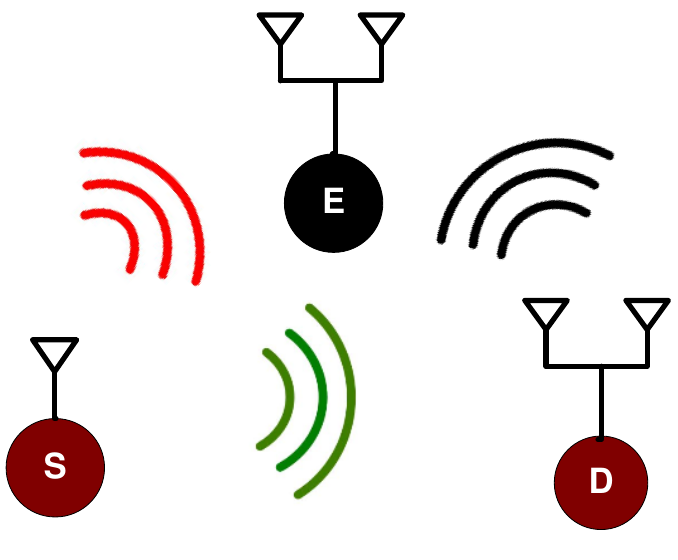}
\end{subfigure}%
\begin{subfigure}
\centering
\includegraphics[width=0.5\textwidth]{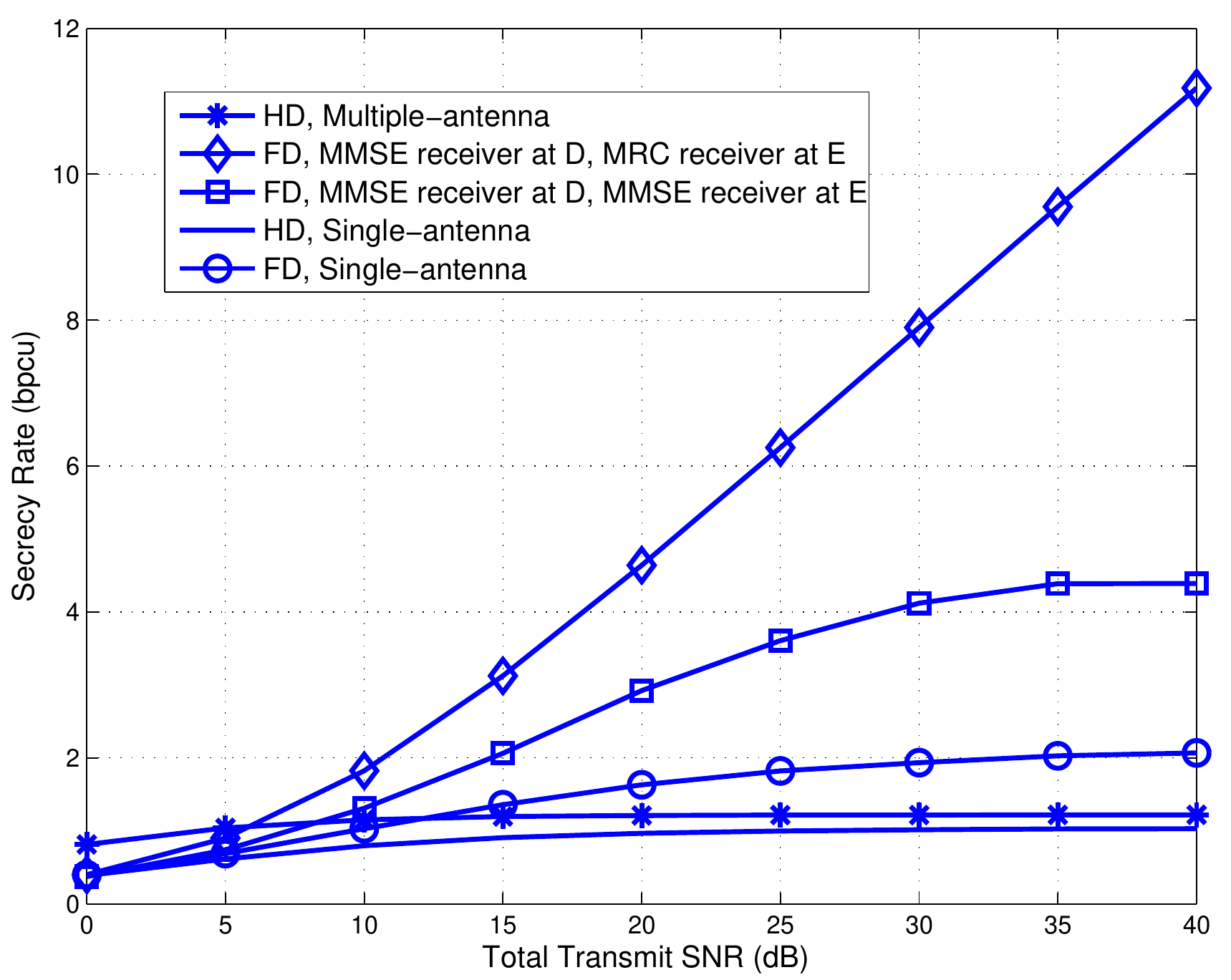}
\end{subfigure}%
\caption{Left:  FD operation at the receiver that creates
self-interference to  improve the secrecy rate. Right: achievable
secrecy rate in bits per channel use (bpcu) vs transmit SNR in
dB.}\label{fig:secrecy}
\end{figure}

 When neither multiple antennas at $\tS$ nor trusted   helpers are available, the system must rely on itself to achieve secure
 communication. To this end,   a self-protection scheme has been proposed that adopts full-duplex (FD) operation at $\tD$ to improve the
secrecy rate \cite{Zheng_TSP}, as shown in Fig. \ref{fig:secrecy}.
More specifically,  an FD receiver is introduced  that
simultaneously receives its data while   transmitting a jamming
signal  to confuse $\tE$. The proposed approach uses intentional
interference at $\tD$ to confuse the eavesdroppers and does not
require external helpers  or data retransmission. Due to the
   FD operation, the receiver experiences a loop
interference (LI) introduced by the transmitted jamming signal. If
$\tD$ has   multiple transmit or receive antennas, it can employ
joint transmit and receive beamforming  for simultaneous signal
detection, suppression and intentional jamming.

In Fig. \ref{fig:secrecy},   the achievable secrecy rate is
evaluated against transmit SNR. We simulate  two cases: i) single
 transmit/receive-antenna receiver and eavesdropper and ii)   the
receiver has two transmit and two receive antennas while the
eavesdropper has four antennas for fairness. For the single-antenna
case, it is seen that the FD scheme outperforms the HD operation for
transmit SNR greater than $10$ dB, and double secrecy rate  is
achieved in the high SNR region. The performance of the HD scheme
saturates when the transmit SNR is higher than $20$ dB. When the
receiver has multiple antennas and the eavesdropper adopts a simple
MRC receiver, the secrecy rate strictly increases with the transmit
SNR and does not saturate at high SNR  as the half-duplex (HD) case.
{  When the eavesdropper is aware of the FD operation at $\tD$ and
adopts the minimum-mean-square-error (MMSE) receiver to mitigate the
jamming signals from $\tD$, the achievable secrecy rate saturates at
a high SNR of 40 dB but is still  significantly higher than the case
with HD receiver.} This reveals the great potential of using
interference at the receiver side to provide self-protection against
eavesdropping.

\section{Conclusions}

In this article, we have introduced radical views on interference in
wireless networks. Traditional interference mitigation techniques
are no longer optimal, and innovative ways of utilizing interference
are emerging. As more aggressive  resource sharing and tighter
cooperation are foreseen in future wireless networks,  interference
management will continue to be a growing challenge.  {\bl
Accordingly, it is essential to further these new perspectives on
interference for more efficient radio resource utilization in
advanced wireless concepts such as large-scale antenna arrays
(massive MIMO),  multicell cooperation, cognitive radio and
heterogeneous networks.  Indeed, the employment of massive MIMO  in
future networks, \bl  allows the mitigation of interference  using
simple linear operations. This way, interference could be
``available'' in the network for other purposes without affecting
its performance; this scenario motivates new services and
applications. In   future cloud radio access networks, baseband
processing will be shifted from BSs to the central baseband unit
pool to jointly process data to and from multicells, and this gives
great opportunities to fully utilize interference. In cognitive
radio, the interference from the secondary user to the primary user
can facilitate RF energy transfer and be tuned into useful signals
if the primary data is known at the secondary user.   Regarding
security in heterogeneous networks, a promising direction is to
study how network interference can be engineered to best benefit
wireless network secrecy.}


\vspace{-5mm}
\begin{thebibliography}{1}
\bibitem{Costa} M. Costa, \textquotedblleft Writing on dirty paper,'' \textit{IEEE Trans. Inf. Theory}, vol. IT-29, pp. 439-441, May 1983.



\bibitem{ElGamal_Kim}
A. El Gamal and Y. H. Kim, \emph{Network Information Theory.}
Cambridge University Press, 2012.

\bibitem{NG11}
{  B. Nazer and M. Gastpar, ``Reliable physical layer network
coding,'' \emph{Proc. IEEE}, vol. 99, pp. 438--460, Mar. 2011.}

\bibitem{Jafar-ia}
S. A. Jafar, ``Interference Alignment  A New Look at Signal
Dimensions in a Communication Network", {\em Foundations and Trends
in Communications and Information Theory}, vol. 7, Issue 1, pp
1-134, 2010.

\bibitem{Jafar14} {  S. A. Jafar, ``Topological Interference Management Through Index Coding,'' \emph{IEEE Trans. on Inf. Theory,} vol. 60 , no. 1, pp. 529-568,  Jan. 2014.}

\bibitem{LTY11}
J. Lee, D. Toumpakaris and W. Yu, ``Interference mitigation via
joint detection,'' \emph{IEEE J. Sel. Area. Comm.}, vol. 29, no. 6,
pp. 1172-1184, Jun. 2011.


\bibitem{cchristos}
C. Masouros, \textquotedblleft Correlation Rotation Linear Precoding
for MIMO Broadcast Channel," \textit{IEEE Trans.  Sig. Process.},
vol. 59, no. 1, pp. 252-262, Jan. 2011.


\bibitem{CI_ComMag} C. Masouros, T. Ratnarajah, M. Sellathurai, C. Papadias, A. Shukla, \textquotedblleft Known Interference in Wireless Communications:
A Limiting factor or a Potential Source of Green Signal Power?,''
\textit{IEEE Comms. Mag., } vol.51, no.10, pp.162-171, Oct. 2013.


\bibitem{Krishnamachari-13}
R. T. Krishnamachari and M. K. Varanasi, ``Interference alignment
under limited feedback for MIMO interference channels,'' {\em IEEE
Trans. Sig. Process.}, vol. 61, no. 15, pp. 3908-3917, Aug. 2013.

\bibitem{Lee_lim10}
N. Lee, J.-B. Lim and J. Chun, ``Degrees of Freedom of the MIMO Y
Channel: Signal Space Alignment for Network Coding," {\em IEEE
Trans.  Inf. Theory}, vol.56, no.7, pp.3332-3342, Jul. 2010.

\bibitem{ZDing_Poor13}
Z. Ding and H. V. Poor, ``A General Framework of Precoding Design
for Multiple Two-Way Relaying Communications," {\em IEEE Trans. Sig.
Process.}, vol.61, no.6, pp.1531-1535, Mar. 2013.



\bibitem{Zhang_PS_TS} R. Zhang and C. K. Ho, ``MIMO broadcasting for simultaneous
wireless information and power transfer,'' {\em IEEE Trans. Wireless
Commun.}, vol. 12, no. 5, pp. 1989-2001, May 2013.




\bibitem{Krikidis_ICC}
 {\bl S. Timotheou, I. Krikidis, G. Zheng, and  B. Ottersten,
``Beamforming for MISO Interference Channels with QoS and RF Energy
Transfer,''   {\em IEEE Trans.   Wireless Commun.}, vol. 13. no. 5,
pp. 2646 - 2658, May 2014. }


\bibitem{Interference_security}
A. Conti, A. Rabbachin, J. Lee, and M. Z. Win, ``Interference
engineering for heterogeneous wireless networks with secrecy,'' in
{\em Proc. Asilomar Conf. Signals, Systems, and Computers}, Pacific
Grove, CA, Nov. 2013, pp. 308-312.

\bibitem{Zheng_TSP}
 G. Zheng, I. Krikidis, J. Li, A. P. Petropulu, and B. Ottersten, ``Improving Physical Layer Secrecy Using Full-Duplex Jamming Receivers'',
  {\em IEEE Trans.  Sig. Process.}, vol. 61, no. 20, pp. 4962-4974 , Oct. 2013.


\end{thebibliography}
 \end{document}